\documentclass[11pt]{article}
\usepackage{amsmath, amsfonts, amssymb}
\usepackage{graphicx}

\begin{document}

\date{}

\title{Observation of Preformed Electron-Hole Cooper Pairs}

\author{
Marijn A. M. Versteegh$^{1}$, A. J. van Lange$^{1}$, H. T. C. Stoof$^{2}$,\\
Jaap I. Dijkhuis$^{1}$\footnote{j.i.dijkhuis@uu.nl} \\
\\
\small $^{1}$Debye Institute for Nanomaterials Science, Utrecht University,\\
 \small Princetonplein 1, 3584 CC Utrecht, The Netherlands \\
  \small $^{2}$Institute for Theoretical Physics, Utrecht University,\\
  \small Leuvenlaan 4, 3584 CE Utrecht, The Netherlands}

\maketitle

\small
Electrons and holes in a semiconductor form hydrogen-atom-like bound states, called excitons. At high electron-hole densities the attractive Coulomb force becomes screened and
excitons can no longer exist. Bardeen-Cooper-Schrieffer theory predicts that at such high densities co-operative many-body effects can at low temperatures induce a bound
state, an electron-hole Cooper pair, comparable to an electron-electron Cooper pair in a superconductor. Here we report the first observation of preformed electron-hole Cooper
pairs in a semiconductor. By measuring stimulated emission from a dense electron-hole gas in ZnO, we have explored both the crossover from the electron-hole plasma to the
preformed Cooper-pair regime, and the crossover from the exciton to the preformed Cooper-pair regime.\\

\normalsize
Superconductivity was successfully explained in 1957 by Bardeen, Cooper, and Schrieffer \cite{bardeen 1957} as the result of Bose-Einstein condensation of many-body
induced bound states of two electrons, called ``Cooper pairs.'' In principle, Bardeen-Cooper-Schrieffer (BCS) theory also allows for many-body induced bound states of other
kinds of fermionic particles. Indeed, in the last decade condensates of Cooper pairs consisting of two fermionic atoms have been observed [2-4]. In 1964, Keldysh and Kopaev predicted the possibility of Cooper pairs of an electron and a hole \cite{keldysh 1964}. In contrast to electron-electron Cooper
pairs, electron-hole Cooper pairs can recombine, and are therefore subject to spontaneous and stimulated emission of photons. Electron-hole Cooper pairs have been
theoretically studied by a number of authors [6-15]. However, measurements of this type of Cooper pairs were never reported, neither of condensed, nor of uncondensed, i.e., preformed pairs.

\begin{figure}
\begin{center}
\includegraphics[width=0.7\textwidth]{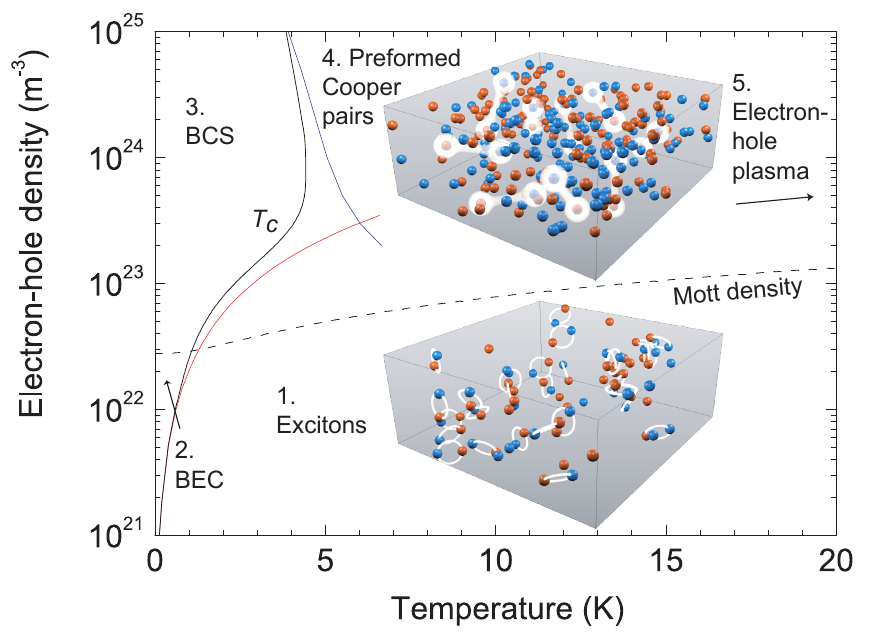}
\caption{\small{(color online). Phase diagram of the electron-hole gas in ZnO. There are five regions: 1. Exciton gas. 2. Excitonic Bose-Einstein condensate. \mbox{3. Electron}-hole BCS state, containing a condensate of electron-hole Cooper pairs. Also called ``non-equilibrium excitonic insulator.'' \mbox{4. Preformed} electron-hole Cooper-pair gas. 5. Electron-hole plasma. The blue line is the calculated mean-field critical temperature $T_C$ in the BCS regime. The red line is the ideal-gas $T_C$ for Bose-Einstein condensation, valid in the BEC regime. Note that the solid black line is the only line of true phase transitions in the phase diagram separating the normal and superfluid phases of the electron-hole gas. In contrast, the dashed line denoting the Mott density represents only a smooth crossover between different regions in the phase diagram. The insets are visualizations of the electron-hole gas in the exciton regime and in the preformed Cooper pair regime.}}
\end{center}
\end{figure}

In this Letter we present the first experimental observation of preformed electron-hole Cooper pairs in a semiconductor. This is achieved by highly exciting a ZnO single
crystal via three-photon absorption at cryogenic temperatures and measuring the light emission. To explain our observations, we show in Fig. 1 the phase diagram of the
electron-hole gas in ZnO. For moderate densities and for temperatures above the critical temperature $T_C$, electrons and holes form an almost ideal gas of excitons (region
1). These excitons undergo Bose-Einstein condensation below $T_C$ (region 2) \cite{blatt 1962, moskalenko 2000}. Bose-Einstein condensation of excitons in a semiconductor has
never been observed. However, impressive results have been obtained in closely related systems. Bose-Einstein condensates (BEC's) of exciton-polaritons have been observed in
semiconductor-based microcavities, where the very small mass of the polaritons facilitates condensation \cite{kasprzak 2006, balili 2007}. In electron-electron bilayers in
high magnetic fields condensation of electron-hole pairs has been detected, where the electrons are in one of the layers and the holes are in the other \cite{kellogg 2004,
tutuc 2004}.

Returning to ZnO, when the electron-hole density of the BEC is increased, the Coulomb forces become gradually screened, leading to a weaker binding of the excitons. Above the
so-called Mott density, the attractive Coulomb force is too weak for excitons to exist. In this high-density, low-temperature regime BCS theory predicts a condensate of a
physically different type of electron-hole bound states: electron-hole Cooper pairs (region 3). The binding of an electron-hole Cooper pair is a combined result of weak
Coulomb attraction and Pauli blocking in the degenerate electron-hole Fermi gases. The energy level of the pairs lies no longer within the band gap, but is located near the
electron-hole Fermi level, which is the energy interval between the electron Fermi level and the hole Fermi level. The condensate of electron-hole Cooper pairs opens up a gap
in the electron-hole pair energy spectrum.

It is interesting to compare the phase diagram of the electron-hole gas with that of a cold gas of fermionic atoms. Where the interaction strength between electrons and holes
can be tuned by varying the electron-hole density, the interaction strength between fermionic atoms can be controlled using a Feshbach resonance \cite{tiesinga 1993}. By
sweeping an external magnetic field across a Feshbach resonance, weakly bound diatomic molecules can be created \cite{regal 2003}, which are the analogue of excitons. Below a
critical temperature these molecules form a BEC \cite{jochim 2003}. Atomic Cooper pairs can be made by tuning the magnetic field to the other side of the Feshbach resonance
\cite{regal 2004}. In particular, the BCS gap \cite{chin 2004} and vortices \cite{zwierlein 2005} have been observed in the atomic BCS superfluid. Recently, it has also been
demonstrated that a temperature range exists above the critical temperature, where uncondensed atomic Cooper pairs are present, accompanied by a pseudogap in the energy
spectrum \cite{gaebler 2010}. Pseudogaps and uncondensed Cooper pairs have also been found in high-$T_C$ superconductors
above the critical temperature, although the relation between those two is more complicated and less understood than for atomic systems \cite{yang 2008, kondo 2011}.

Returning to our phase diagram, we anticipate that such a regime of uncondensed Cooper pairs also exists for the electron-hole gas (region 4). In agreement with common
practice, we call the uncondensed pairs ``preformed electron-hole Cooper pairs,'' while the term ``electron-hole Cooper pair'' refers to the condensed state. When the temperature is
increased above a crossover temperature $T^*>T_C$, the preformed electron-hole Cooper pairs dissociate, resulting in an electron-hole plasma (region 5). Lowering the density
below the Mott density leads to the formation of excitons again.

The Mott density in Fig. 1 we calculated from the condition that at the Mott density the screening length equals the exciton Bohr radius in the unscreened case \cite{haug
2004}. For the critical temperature in the BEC regime we used the well-known ideal gas result \cite{kittel 1980}. To obtain $T_C$ in the BCS regime, we solved the BCS gap
equation in mean-field approximation. Details of these calculations are given in the Supplemental Information. The most important result is that the temperatures and densities
at which preformed electron-hole Cooper pairs form in ZnO seem to be well accessible by experiment. Even the superfluid phase might be reachable.

To explore the physics of the phase diagram, we measured the light emission from a highly excited ZnO single crystal. The crystal was 500 $\mu$m thick and oriented in the
[0001] direction. ZnO was chosen for our study for several reasons. Firstly because its strong electron-hole Coulomb pairing, apparent in the large exciton binding energy of
60 meV, leads to a high $T_C$. Secondly because its direct band gap results in strong light emission from the crystal. Thirdly because its simple band structure is unfavorable
for the formation of an electron-hole liquid or electron-hole droplets. In Si and Ge, the many-valley structure of the conduction bands, the fourfold degeneracy of the valence
bands, and the anisotropy of the corresponding masses lead to the formation of electron-hole liquids and electron-hole droplets, and thus prevent the formation of
electron-hole Cooper pairs \cite{brinkman 1973, moskalenko 2000}.

The fourth reason for choosing ZnO is that it is possible to create high electron-hole densities in the bulk of a ZnO crystal via three-photon absorption of high-intensity
infrared laser pulses. Low temperature luminescence from highly excited ZnO was studied before in the case of direct excitation by ultraviolet light pulses \cite{priller
2004}, where the short penetration depth limits the excitation layer to a thickness of 50 nm. To avoid emission related to surface impurities and explore the bulk physics, we
employed the long penetration depth of infrared light, and excited our crystal slab via three-photon absorption of high-intensity 160-fs 800-nm pulses from an amplified
Ti:sapphire laser. An additional feature of this approach is that spontaneous emission by a preformed electron-hole Cooper pair triggers stimulated emission from other
preformed electron-hole Cooper pairs as the emitted photon traverses the 500 $\mu$m long excited zone. This amplified spontaneous emission (ASE) is relatively easy to measure
and recognize. The electron-hole density in our experiment is controlled via the intensity of the excitation pulse, and calculated from the known three-photon absorption
coefficient \cite{versteegh 2011}.

\begin{figure}
\begin{center}
\includegraphics[width=0.7\textwidth]{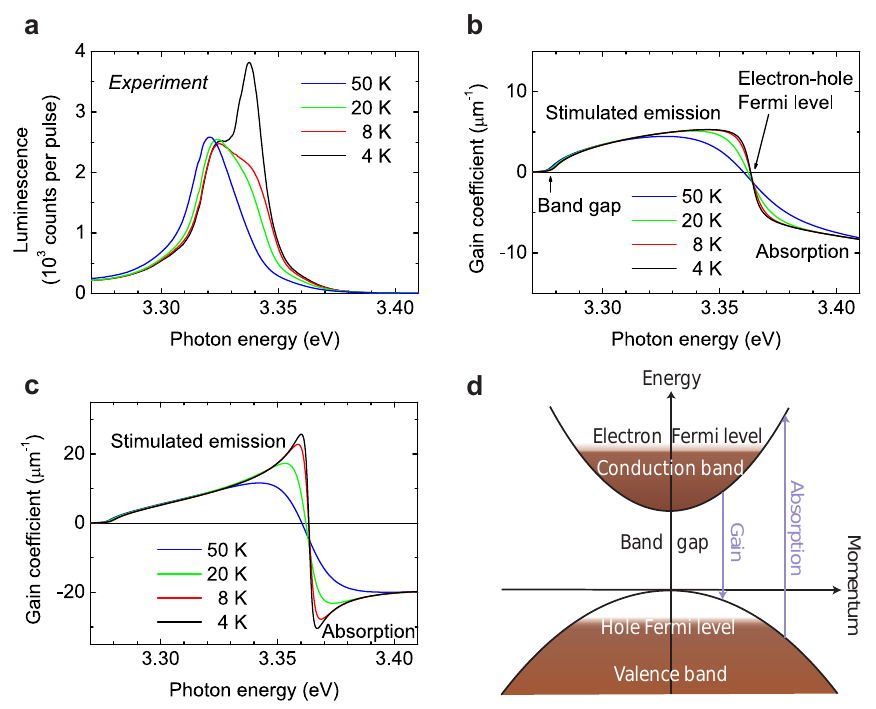}
\caption{\small{(color online). Crossover from the electron-hole plasma regime to the preformed electron-hole Cooper-pair regime for decreasing temperature. The electron-hole density is $n = 9.2\times10^{24}$ m$^{-3}$. (a) Measured emission spectra of the ZnO crystal. At \mbox{$T = 50$ K} there is only emission from an electron-hole plasma. For decreasing temperature, gain (amplified spontaneous emission) from preformed electron-hole Cooper pairs appears. (b) Theoretical gain spectra without Coulomb forces between electrons and holes. (c) Theoretical gain spectra taking into account Coulomb forces, showing the appearance of preformed electron-hole Cooper-pair peaks. (d) Scheme of the band structure and the degenerate electron-hole gas. For our calculations we used a two-band model including spin degeneracy, which at low temperatures is a good approximation for densities below about \mbox{$10^{25}$ m$^{-3}$.}}}
\end{center}
\end{figure}

Measured emission spectra at high electron-hole density ($n=9.2\times10^{24}$ m$^{-3}$) for decreasing temperature are shown in Fig. 2a. At $T = 50$ K we measure mainly
spontaneous emission from an electron-hole plasma. When the ZnO crystal is cooled to 4 K, a strong new peak emerges. On the basis of the calculated phase diagram (Fig. 1), it
can be expected that at these temperatures the electron-hole gas has entered the preformed electron-hole Cooper-pair regime.

To find out whether this observed new peak is related to preformed electron-hole Cooper pairs, we calculated gain spectra of the electron-hole gas in ZnO, using the many-body
theory explained in Ref. \cite{versteegh 2011}. In this theory, the interaction between the carriers is described by the Yukawa potential and the optical spectra of ZnO are
calculated by solving the statically screened Bethe-Salpeter ladder equation. This theory has been experimentally verified at room temperature for bulk ZnO \cite{versteegh
2011}. Further details of our calculations are given in the Supplemental Information. Figure 2b shows the theoretical gain spectra without interactions between the electrons
and holes. In the spectra of Fig. 2c the Coulomb forces between electrons and holes have been taken into account. In both cases there is a spectral domain where gain occurs
and a domain where absorption prevails, as can be understood from Fig. 2d. The crucial difference is that for decreasing temperature two peaks appear when Coulomb forces are
included: a gain peak just below the electron-hole Fermi level, and an absorption peak just above this level. In the theory these peaks arise from fluctuations in the
electron-hole pairing fields, showing that they are due to preformed electron-hole Cooper pairs, the precursor of the electron-hole Cooper-pair condensate. In the two-dimensional case the same physical interpretation of these peaks was obtained in Ref. \cite{schmitt-rink 1986}.

Comparison between theory and experimental data suggests that the observed new peak is due to stimulated emission from preformed electron-hole Cooper pairs. Indeed, additional
experimental results (see Supplemental Information) at $n=1.9\times10^{23}$ m$^{-3}$ show that all emission peaks have longitudinal-optical phonon replica, except the new
peak, which is evidence that the new peak is not just spontaneous emission like the other peaks, but is associated with gain. We therefore attribute this peak to ASE from
preformed electron-hole Cooper pairs. The difference in spectral position between the measured new peak of Fig. 2a and the calculated peak of Fig. 2c originates from limited
precision in the calculation of the band-gap renormalization.

\begin{figure*}
\begin{center}
\includegraphics[width=\textwidth]{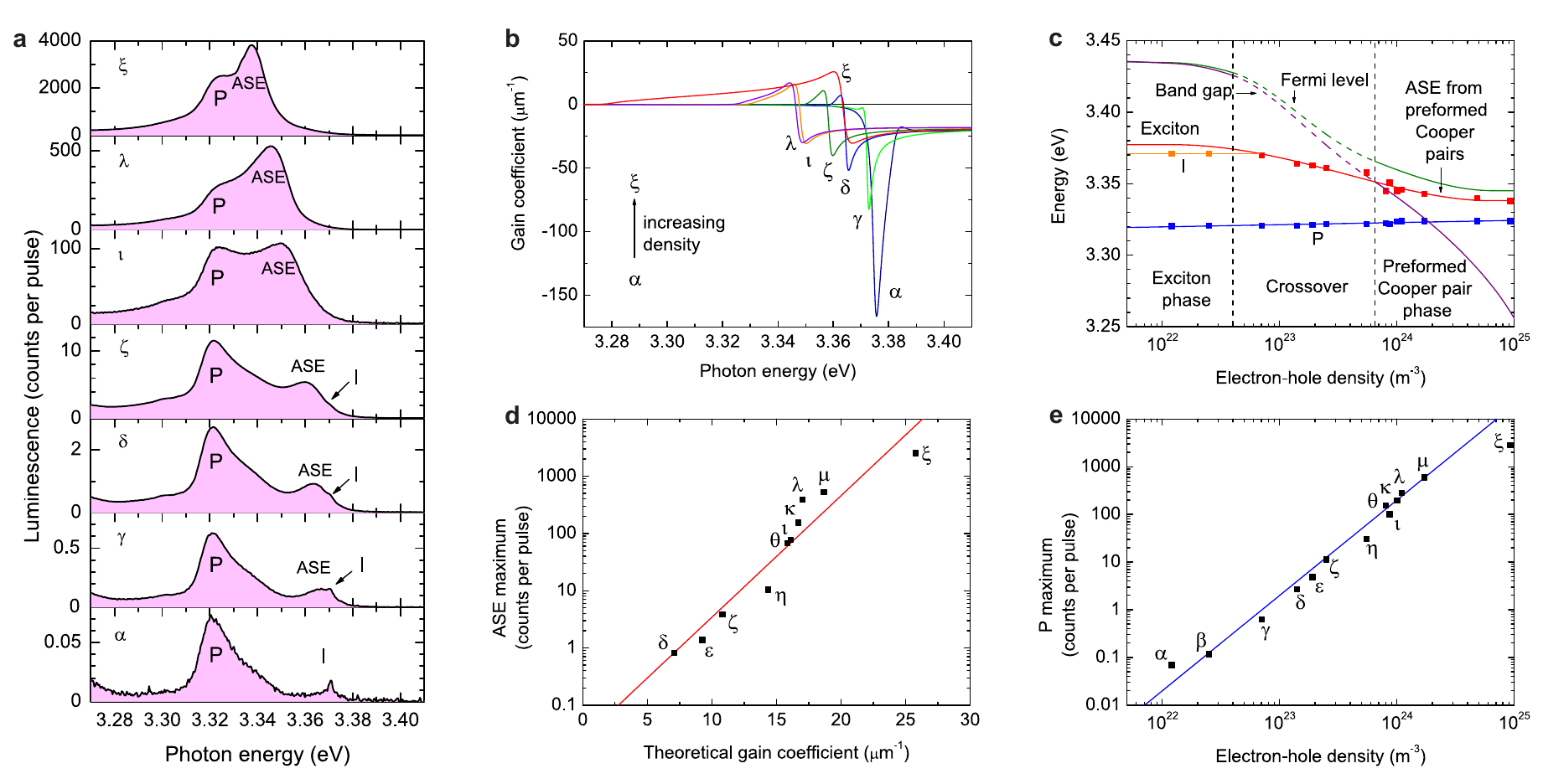}
\caption{\small{(color online). Crossover from the exciton regime to the preformed Cooper-pair regime for increasing electron-hole density. $T = 4$ K. Greek symbols indicate
electron-hole densities (in m$^{-3}$): $\alpha$ $1.2\times10^{22}$, \mbox{$\beta$ $2.5\times10^{22}$,} \mbox{$\gamma$ $7.0\times10^{22}$,} \mbox{$\delta$ $1.4\times10^{23}$,}
$\varepsilon$ $1.9\times10^{23}$, $\zeta$ $2.5\times10^{23}$, \mbox{$\eta$ $5.5\times10^{23}$,} \mbox{$\theta$ $8.0\times10^{23}$,} \mbox{$\iota$ $8.6\times10^{23}$,} $\kappa$
$1.0\times10^{24}$, $\lambda$ $1.1\times10^{24}$, \mbox{$\mu$ $1.7\times10^{24}$,} \mbox{$\xi$ $9.2\times10^{24}$.} (a) Measured emission spectra. At low densities we observe
spontaneous emission: the P peak and impurity-bound exciton emission (I). For increasing density gain (ASE) from preformed electron-hole Cooper pairs appears. (b) Theoretical
gain spectra showing the appearance of a preformed Cooper-pair gain peak. Note that the P peak is absent here, since the theory does not incorporate pair-pair collisions. (c)
Squares indicate measured emission energies. The band-gap renormalization and the electron-hole Fermi level were derived from these. We have drawn dashed lines in the
crossover, since a theory is presently lacking for this regime. (d) Measured ASE maxima versus theoretical gain coefficient maxima, showing an exponential relation. (e)
Measured P peak maxima versus electron-hole density, showing a quadratic dependence.}}
\end{center}
\end{figure*}

In Fig. 3 we explore the crossover from the exciton regime to the preformed Cooper-pair regime at $T = 4$ K by tuning the electron-hole density. For clarity, a subset of
measured emission spectra is shown in Fig. 3a. The complete data set is presented in the Supplemental Information. At the lowest density the well-known spontaneous emission
spectrum is observed: The highest peak is the so-called P peak, resulting from inelastic scattering of two excitons, where one recombines and the other ionizes \cite{hvam
1973}. The emission at 3.371 eV results from recombination of excitons bound to aluminium impurities \cite{meyer 2007}. This emission spectrum shows that the electron-hole gas
is in the exciton regime, as could be expected from the calculated phase diagram (Fig. 1). When the density is increased to $7\times10^{22}$ m$^{-3}$, a new peak appears, next
to the impurity peak. For increasing density this peak grows very fast, even faster than the P peak, and the spectral distance between the P peak and this new peak decreases.

To examine the origin of this new peak, we calculated the gain spectra at $T = 4$ K, shown in Fig. 3b, using the same theory as for Fig. 2c. At low densities, these spectra
show excitonic absorption and no gain, in agreement with our interpretation of the observed emission as spontaneous emission from a exciton gas. For increasing electron-hole
density, the spectra gradually evolve into a double-peak structure, characteristic for (preformed) electron-hole Cooper pairs \cite{schmitt-rink 1986}. This evolution is in
agreement with the phase diagram (Fig. 1). Gain appears in the theoretical spectra at $8\times10^{22}$ m$^{-3}$, very close to the threshold for the new peak. We conclude that
the new peak originates from ASE from preformed electron-hole Cooper pairs, or, for densities not far above the Mott density, from bound electron-hole pairs in the crossover
between excitons and preformed electron-hole pairs. Note that the impurity emission is still visible, separated from the new peak, showing that these contributions to the
emission spectrum have different origins.

Interestingly, the P peak persists even at high carrier densities where excitons cannot exist. In the preformed Cooper-pair regime we interpret the P peak in an analogous way
as for excitons, namely as due to inelastic scattering of two preformed electron-hole Cooper pairs, in which one pair recombines and the other breaks up. This interpretation
is in line with calculations by Inagaki and Aihara \cite{inagaki 2002} in the BEC-BCS crossover. Consequently, the energy separation between the P maximum and the ASE maximum
directly measures the binding energy of the preformed electron-hole Cooper pairs.

The spectral positions of the measured peaks are analyzed in Fig. 3c. Here we have indicated the exciton region and the preformed Cooper-pair region, connected by a crossover
region above $n = 4\times10^{22}$ m$^{-3}$, the calculated Mott density at $T = 4$ K. In the crossover, the exciton energy level gradually develops into the preformed
electron-hole Cooper-pair level, which is equal to the spectral position of the gain peak. The diminishing separation between the P and the ASE peak indicates the reduction of
the electron-hole-pair binding energy by screening.

It is important to realize that band-gap renormalization takes place: The band gap narrows for increasing density due to exchange and correlation effects. From the binding
energies determined from the measured P-ASE peak separation we derive in the following way the values of the density-dependent band gap and the electron-hole Fermi level. In
the exciton regime the band gap is of course located one exciton binding energy above the exciton energy level. In the preformed Cooper-pair regime we do not have such direct
information about the value of the band gap. However, we expect a pseudogap around the electron-hole Fermi level equal to the preformed Cooper-pair binding energy. Therefore,
in this regime the Fermi level lies only half of the binding energy above the preformed Cooper-pair level. Ideal gas chemical potentials yield the separation between the band
gap and the Fermi level. Based on this argumentation we can draw the band gap and the Fermi level in both the exciton region and in the preformed Cooper-pair region. For the
crossover a theory is presently lacking. Therefore we have drawn dashed lines in this regime. The rather good agreement between the determined band-gap renormalization and
curves from the literature (Supplemental Information) is another evidence for the correctness of our theory and our interpretation of the experimental data.

In Fig. 3d measured ASE maxima are plotted against the theoretical maximum gain coefficients at the electron-hole densities realized in the experiment. We find an exponential
relation, what is to be expected for ASE. The P peak height depends quadratically on the carrier density (Fig. 3e). This result confirms our interpretation of the P peak as
the result of scattering of two excitons or two preformed Cooper pairs, depending on the carrier density.

Our calculations of the critical temperature in Fig. 1 indicate that also the BCS state should be experimentally accessible. It is still an open question how this state could
be detected. By using a microcavity setup, as in Refs. \cite{kasprzak 2006, balili 2007}, Bose-Einstein condensation of two-dimensional electron-hole Cooper-pair polaritons
may occur at even higher temperatures. We expect that our observation of preformed electron-hole Cooper pairs in a semiconductor will contribute to the understanding of
high-temperature superconductivity, exciton-polariton condensates, and excitonic condensation in electron-hole bilayers and bilayer quantum Hall systems.

We thank C. R. de Kok and P. Jurrius for technical support, J. M. Vogels, E. E. van Faassen, D. van Oosten, P. van der Straten, and J. Lipfert for discussions and reviewing of
our manuscript and D. van Oosten for drawing the insets of Fig. 1.

\end{document}


\date{}

\title{SUPPLEMENTAL INFORMATION\\
Observation of Preformed Electron-Hole Cooper Pairs}

\author{
Marijn A. M. Versteegh$^{1}$, A. J. van Lange$^{1}$, H. T. C. Stoof$^{2}$,\\
Jaap I. Dijkhuis$^{1}$\footnote{j.i.dijkhuis@uu.nl} \\
\\
\small $^{1}$Debye Institute for Nanomaterials Science, Utrecht University,\\
 \small Princetonplein 1, 3584 CC Utrecht, The Netherlands \\
  \small $^{2}$Institute for Theoretical Physics, Utrecht University,\\
  \small Leuvenlaan 4, 3584 CE Utrecht, The Netherlands} \maketitle

\section{Theory of the electron-hole phase diagram}\label{theory phase diagram}

In this Section we explain the calculation of the phase diagram of the electron-hole gas in ZnO, Fig. 1. First we discuss how we calculated the Mott density, which marks the crossover between the exciton region and the preformed Cooper pair region, or the electron-hole plasma region. Then we explain the calculation of the critical temperature, below which the electron-gas forms a BEC or a BCS superfluid.

\subsection{Mott density}

The Mott density is the density above which excitons do not exist due to screening of the Coulomb attractive interaction between an electron and a hole. We describe this screened interaction by the Yukawa potential
\begin{equation}\label{yukawa real space}
V_s (r,n,T)=\frac{-e^2}{4\pi \varepsilon_0 \varepsilon_r r}e^{-r/\lambda_s (n,T)},
\end{equation}
where $r$ is the distance between the electron and the hole, $-e$ is the electron charge, $\varepsilon_0$ is the vacuum permittivity, $\varepsilon_r$ is the relative dielectric constant of ZnO, and $\lambda_s (n,T)$ is the screening length, which depends on the electron-hole density $n$ and the temperature $T$. An excitonic bound state exists in this potential when $\lambda_s (n,T)>a_0$, where $a_0$ is the unscreened exciton Bohr radius \cite{haug 2004}. Therefore, at the Mott density $\lambda_s (n,T)=a_0$. In ZnO $a_0=4\pi \hbar^2 \varepsilon_0 \varepsilon_r/(e^2 m_r)=1.83$ nm, with $m_r=0.19\,m_0$ the reduced mass of the exciton and $m_0$ is the bare electron mass.

The screening length is calculated in the random-phase approximation (RPA) from the ideal gas chemical potentials of the electron and hole gases, $\mu_e(n,T)$ and $\mu_h(n,T)$, respectively. For details we refer to Ref. \cite{versteegh 2011}. The result for the Mott density is shown in Fig. 1. We see that the Mott density rises for increasing temperature, reflecting the reduced screening efficiency by the electrons and holes due to their thermal motion.

\subsection{Superfluid critical temperature}

We computed the superfluid critical temperature $T_C$ as a function of density both in the BEC regime and in the BCS regime. In the BEC regime the critical temperature is given by the well-known ideal gas result \cite{kittel 1980}
\begin{equation}\label{BEC TC}
T_C(n)=\frac{2\pi\hbar^2}{m_e+m_h}\Big(\frac{n}{4\times2.612}\Big)^{2/3},
\end{equation}
where $m_e=0.28\,m_0$ and $m_h=0.59\,m_0$ are the effective electron and hole masses, respectively \cite{button 1972, hummer 1973}. The factor 4, multiplying 2.612, is included because of the four possible spin states of the exciton \cite{nozieres 1982}. The effective interaction between two excitons is at present unknown, but is most likely repulsive, corresponding to a positive scattering length. In that case it would lead to a small upward shift of the critical temperature proportional to $n^{1/3}$. This effect is, however, neglected here.

In the BCS regime we calculated $T_C$ in mean-field approximation by solving the linearized BCS gap equation \cite{conti 1998, stoof 2009}
\begin{equation}\label{gap normal state}
\Delta_{\mathbf{k}}(n,T)\!=\!\!\int \!\!\frac{\mathrm{d}\mathbf{k'}}{(2\pi)^3}V_{s,|\mathbf{k}-\mathbf{k'}|}(n,T)\frac{1-f_{\mathbf{k'},e}(n,T)-f_{\mathbf{k'},h}(n,T)} {\mu_e(n,T)\!-\!\varepsilon_{\mathbf{k'},e}\!+\!\mu_h(n,T)\!-\!\varepsilon_{\mathbf{k'},h}}\Delta_{\mathbf{k'}}(n,T),
\end{equation}
where $\Delta_{\mathbf{k}}(n,T)$ is the momentum-dependent BCS order parameter or gap, and
\begin{equation}\label{yukawa momentum space}
V_{s,|\mathbf{k}-\mathbf{k'}|}(n,T)=\frac{-e^2}{\varepsilon_0 \varepsilon_r (\lambda_s^{-2}(n,T)+|\mathbf{k}-\mathbf{k'}|^2)}
\end{equation}
is the Yukawa potential in momentum space, i.e., the Fourier transform of Eq. (\ref{yukawa real space}). Furthermore, $\varepsilon_{\mathbf{k},i}$ are the single-particle energies $\varepsilon_{\mathbf{k},i}=\varepsilon_{k,i}=\hbar^2k^2/(2m_i)$, where $i$ stands for $e$, electron, or $h$, hole, and $k=|\mathbf{k}|$. Note that we use a momentum-dependent interaction, not a point interaction or separable pseudopotential as in standard BCS theory. Finally,
\begin{equation}
f_{\mathbf{k},i}(n,T)=\frac{1}{e^{[\varepsilon_{\mathbf{k},i}-\mu_i(n,T)]/(k_B T)}+1}
\end{equation}
are the Fermi-Dirac distributions of the electron and hole gases.

Eq. (\ref{gap normal state}) is valid for $T\geq T_C$. For $T> T_C$ the BCS gap equation has only one solution, namely $\Delta_{\mathbf{k}}=0$. This solution corresponds to the normal phase. Below $T_C$ also a solution $\Delta_{\mathbf{k}}\neq0$ is possible, corresponding to the superfluid phase. Since the normal to BCS phase transition is a second-order phase transition in the balanced case, the order parameter $\Delta_{\mathbf{k}}$ continuously increases from zero for $T>T_C$ to its maximum at $T=0$. Precisely at $T_C$ the order parameter becomes nonzero.

\renewcommand{\thefigure}{S\arabic{figure}}
\setcounter{figure}{0}

\begin{figure}
\begin{center}
\includegraphics[width=0.8\textwidth]{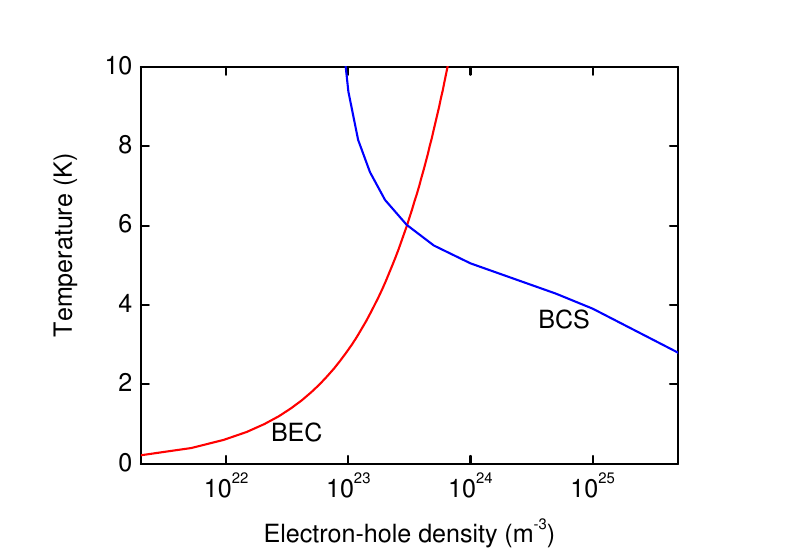}
\caption{Critical temperature. $T_C$ calculated in the BEC regime (Eq. \ref{BEC TC}) and $T_C$ calculated in the BCS regime by solving the BCS equation (Eq. \ref{gap normal state}).\label{7figS1}}
\end{center}
\end{figure}

In order to find the highest temperature at which Eq. (\ref{gap normal state}) has a nonzero solution, we use that fact that for an $s$-wave order parameter we can replace $V_{s,|\mathbf{k}-\mathbf{k'}|}$ by its angle average, depending only on the lengths $k$ and $k'$ as
\begin{equation}
\overline{V}_{s,k,k'}=\frac{1}{2}\int_{0}^{\pi}\textrm{d}\theta\; \sin\theta\; V_{s,|\mathbf{k}-\mathbf{k'}|}
=\frac{-e^{2}}{4\varepsilon_{0}\varepsilon_{r}kk'}\ln\Big[\frac{(k^{2}+k'^{2}+2kk')\lambda_{s}^{2}+1}{(k^{2}+k'^{2}-2kk')\lambda_{s}^{2}+1}\Big].
\end{equation}
In this way the gap equation is written in the form of a homogeneous Fredholm integral equation of the second kind. Such equations are generally not analytically solvable. In order to obtain a numerical solution, we replace the integral by a finite sum. The discretization is preformed following Simpson's rule and leads to the simple eigenvalue problem
\begin{equation}\label{eigenvalue}
\overrightarrow{\Delta}(n,T)=\overrightarrow{\overrightarrow{\mathrm{U}}}(n,T)\cdot\overrightarrow{\Delta}(n,T).
\end{equation}
The gap has become a vector $\overrightarrow{\Delta}(n,T)$ with length $p$, where $p$ is the number of steps in the discretization. The quantity $\overrightarrow{\overrightarrow{\mathrm{U}}}(n,T)$ is a $p\times p$ matrix, whose elements are given by
\begin{equation}
\mathrm{U}_{ij}(n,T)=w_j\frac{1}{2\pi^2}k_j^2 \frac{1-f_{k_j,e}(n,T)-f_{k_j,h}(n,T)}{\mu_e(n,T)-\varepsilon_{k_j,e}+\mu_h(n,T)-\varepsilon_{k_j,h}}\overline{V}_{s,k_i,k_j}(n,T),
\end{equation}
where $w_j$ are the Simpson weights for each discretization point. For each density $n$ we approach $T_C$ from above and search for a nontrivial solution of Eq. (\ref{eigenvalue}). The critical temperature $T_C$ is the temperature at which $\overrightarrow{\overrightarrow{\mathrm{U}}}(n,T)$ has eigenvalue 1. The results for the critical temperature in the BEC and in the BCS regime are shown in Fig. 1 and in Fig. S1.

\subsection{Two-band model}\label{two-band model}

For all calculations we used an isotropic parabolic two-band model, including spin degeneracy. ZnO has one twofold spin-degenerate conduction band and three twofold spin-degenerate valence bands, called A, B, and C. The three valence bands are split as a result of spin-orbit coupling and the hexagonal crystal field. The AB splitting equals 10 meV, the AC splitting 44 meV \cite{lambrecht 2002}. In our model only the conduction band and the the highest valence band, the A band, are taken into account.

At low temperatures and densities where the hole Fermi energy is smaller than the AB splitting this is a good approximation, since almost all holes are in the A band. According to the hole masses determined by H\"{u}mmer \cite{hummer 1973}, the hole Fermi energy equals 10 meV at $n=2\times10^{24}$ m$^{-3}$. According to the nonisotropic energy-dependent hole masses calculated by Lambrecht \textit{et al.} \cite{lambrecht 2002}, however, this is only the case at $n=8\times10^{24}$ m$^{-3}$. For the calculations of the Mott density and of the critical temperature in the BEC regime inclusion of the B and C bands in the calculation would not make a difference. For the critical temperature in the BCS regime an effect would show up at high densities where the occupancy in the B band is sufficiently high that it significantly affects the hole Fermi level. When that happens, a distinction has to be made between the BCS order parameter related to correlations between the A band and the conduction band and the BCS order parameter related to correlations between the B band and the conduction band. Because of the presence of density imbalance, the physics changes and exotic phases like the Sarma phases and the FFLO phase could possibly show up.

\section{Calculation of the gain spectra}\label{theory gain}

The theoretical gain spectra shown in Figs. 2b, 2c, and 3b were calculated using the many-body theory explained in Ref. \cite{versteegh 2011}. This theory has been
experimentally verified at room temperature for bulk ZnO \cite{versteegh 2011}.

In this theory, the interaction between the carriers is described by the Yukawa potential in Eqs. (\ref{yukawa real space}) and (\ref{yukawa momentum space}). The optical
spectra of ZnO are calculated by deriving and solving the statically screened Bethe-Salpeter ladder equation. Thus, we have obtained the gain spectra in the exciton regime, in
the preformed Cooper-pair regime, in the electron-hole plasma regime, and in the crossovers between those regimes. The peaks in Fig. 2c arise due to fluctuations in the
pairing fields $\Delta$, showing that they are really due to (preformed) electron-hole Cooper pairs. Importantly, the expectation value of the $\Delta$-fields, i.e., the BCS
order parameter, is set to zero in Ref. \cite{versteegh 2011}. Therefore, the calculated gain spectra are only valid above $T_C$. Note that according to our calculations (Fig.
1) our measurements were performed above $T_C$. The influence of the fluctuations in the $\Delta$-fields on the self energies of the electrons and holes is neglected in this
theory. As a result of this, the single-particle density of states would not show pseudogaps in our approximation.

The optical spectra in Ref. \cite{versteegh 2011} were calculated for $\mathbf{E}\perp \mathbf{c}$. In our experiment, the axis of the excited cylinder is parallel to the \textit{c}-axis. The amplified spontaneous emission therefore mainly has polarization $\mathbf{E}\perp \mathbf{c}$. Hence, the expressions given in Ref. \cite{versteegh 2011} can also be used in the present case.

For the gain spectra of Figs. 2c and 3b we have performed the same calculation as for the room-temperature case in Ref. \cite{versteegh 2011}, but with the following
low-temperature parameters for the nonrenormalized band gap \cite{hauschild 2006} and damping: $E_{G,0}(4\textrm{K})=E_{G,0}(8\textrm{K})=3.437$ eV,
$E_{G,0}(20\textrm{K})=3.436$ eV, $E_{G,0}(50\textrm{K})=3.434$ eV and $\hbar \gamma_0=2.0$ meV. For the step size $s$ it was necessary to take $5\times10^6$ m$^{-1}$ instead
of $5\times10^7$ m$^{-1}$. Thus, the matrices to be inverted were $501\times501$ instead of $51\times51$. All other parameters were the same as for the room-temperature
calculation.

The spectra shown in Fig. 2b are mean-field spectra. They have been calculated from Eq. (A41) of Ref. \cite{versteegh 2011} with the low-temperature parameters as specified
above. Note that the mean-field susceptibility, Eq. (A41), can be obtained from the RPA Bethe-Salpeter susceptibility, Eq. (15), by putting $V_{s,|\mathbf{k}-\mathbf{k'}|}=0$,
i.e., by ignoring the Coulomb forces between electrons and holes. In all theoretical spectra band-gap renormalization is included, as calculated from the phenomenological
formula of B\'{a}nyai and Koch \cite{haug 2004, banyai 1986}.

As explained in Sec. \ref{two-band model}, we used a two-band model, which is a good approximation at low temperatures and low densities. At densities considerably higher than $2\times10^{24}$ m$^{-3}$ or $8\times10^{24}$ m$^{-3}$, depending on which hole masses are most appropriate in practice, occupancy of states in the B valence band, and ultimately also in the C valence band, would reduce preformed Cooper-pair gain and shift the gain maximum to lower energies.

\pagebreak

\section{Supplemental Figures}

\begin{figure}[!h]
\begin{center}
\includegraphics[width=0.9\textwidth]{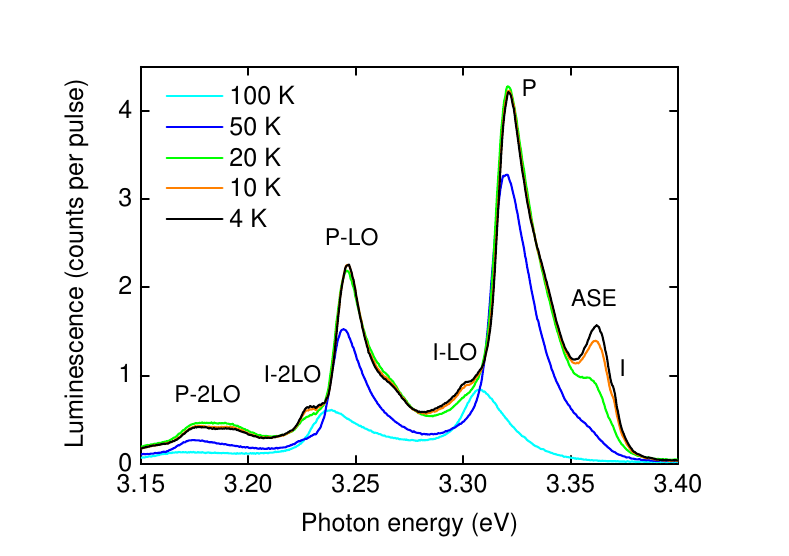}
\caption{Measured emission spectra from the ZnO crystal at electron-hole density $n=1.9\times10^{23}$ m$^{-3}$. For decreasing temperature the electron-hole gas makes a transition from the electron-hole plasma to the preformed Cooper-pair regime. The P emission, indicated by P, and the spontaneous emission from excitons bound to aluminum impurities, indicated by I, have first and second phonon replica: spontaneous emission causing the P and I peaks also occurs under simultaneous emission of one or two longitudinal optical (LO) phonons, leading to emission peaks at lower photon energies. These LO phonons have an energy of 72 meV. The ASE peak appearing for decreasing temperature is caused by spontaneous emission from electron-hole pairs in the crossover from excitons to preformed Cooper pairs, amplified by stimulated emission from such pairs. This ASE peak does not have phonon replica, demonstrating that it is not just spontaneous emission like the other peaks. The general shift to higher photon energies for lower temperatures reflects the temperature-dependence of the band gap \cite{hauschild 2006}.}
\end{center}
\end{figure}

\begin{figure}
\begin{center}
\includegraphics[width=\textwidth]{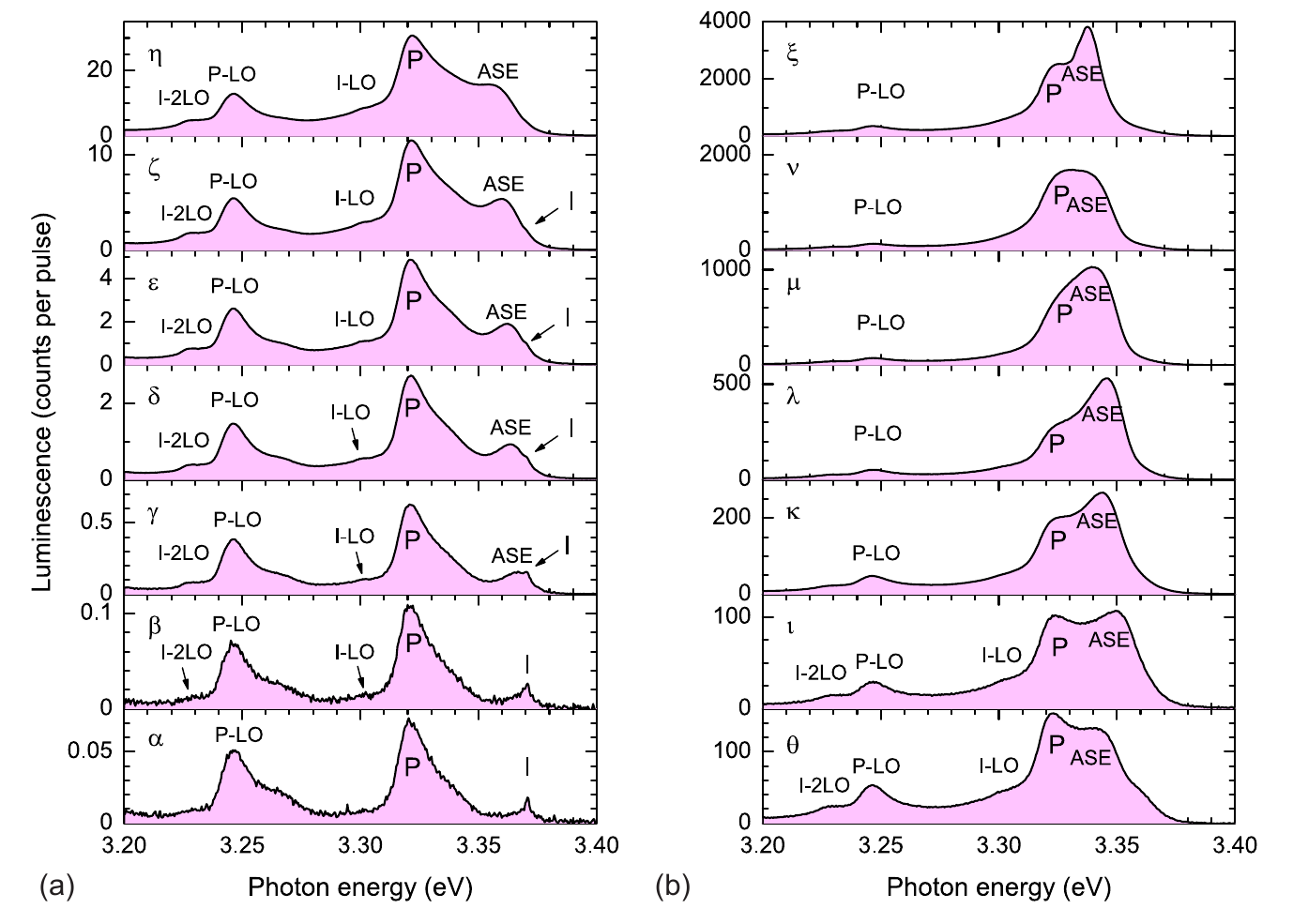}
\caption{Full experimental data set corresponding to Fig. 3, showing the crossover from the exciton regime to the preformed Cooper pair regime at $T = 4$ K. Greek symbols indicate electron-hole densities (in m$^{-3}$): (a) $\alpha$ $1.2\times10^{22}$, \mbox{$\beta$ $2.5\times10^{22}$,} \mbox{$\gamma$ $7.0\times10^{22}$,} $\delta$ $1.4\times10^{23}$, $\varepsilon$ $1.9\times10^{23}$, \mbox{$\zeta$ $2.5\times10^{23}$,} $\eta$ $5.5\times10^{23}$. (b) \mbox{$\theta$ $8.0\times10^{23}$,} $\iota$ $8.6\times10^{23}$, $\kappa$ $1.0\times10^{24}$, $\lambda$ $1.1\times10^{24}$, \mbox{$\mu$ $1.7\times10^{24}$,} $\nu$ $4.8\times10^{24}$, \mbox{$\xi$ $9.2\times10^{24}$.} Indicated are the amplified spontaneous emission (ASE) from preformed Cooper pairs, or from electron-hole pairs in the crossover from excitons to preformed Cooper pairs, the P emission (P), the emission from excitons bound to aluminum impurities (I), the first longitudinal optical phonon replica of the P peak (P-LO), and the first (I-LO) and second \mbox{(I-2LO)} phonon replica of the I peak.}
\end{center}
\end{figure}

\begin{figure}
\begin{center}
\includegraphics[width=0.8\textwidth]{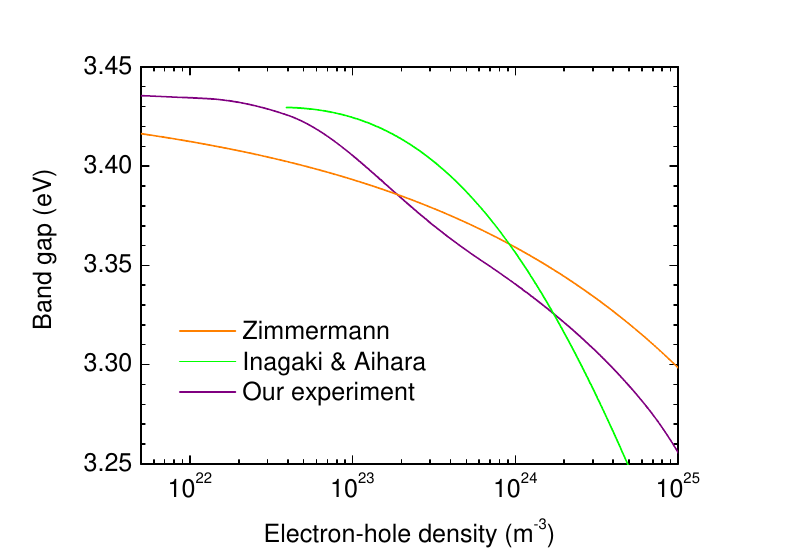}
\caption{Band-gap renormalization. Band gap versus electron-hole density at $T=4$ K as derived from our experimental results (Fig. 3c) compared with two curves from the literature: Zimmermann \cite{zimmermann 1988}, calculated at $T=4$ K and Inagaki and Aihara \cite{inagaki 2002}, calculated at $T=0$.\label{7figBGR}}
\end{center}
\end{figure}

\pagebreak